\documentstyle[apjfonts,emulateapj]{article}

\begin{document}

\title{Jets in GRBs}
\author{Re'em Sari\altaffilmark{}}
\affil{Theoretical Astrophysics, 130-33, California Institute of Technology,
        \\Pasadena, CA 91125}
\author{Tsvi Piran\altaffilmark{1,2} and J. P. Halpern}
\affil{Columbia Astrophysics Laboratory, Columbia University,
        550 West 120th Street,\\New York, NY 10027}

\altaffiltext{1}{Racah Institute of Physics, The Hebrew University, Jerusalem
91904, Israel.}
\altaffiltext{2}{Department of Physics, New York University, New York, NY
10003.}

\begin{abstract}
In several GRBs afterglows, rapid temporal decay is observed which is
inconsistent with spherical (isotropic) blast-wave models. In particular,
GRB~980519 had the most rapidly fading of the well-documented GRB
afterglows, with $t^{-2.05\pm 0.04}$ in optical as well as in X-rays. We
show that such temporal decay is more consistent with the evolution of a jet
after it slows down and spreads laterally, for which $t^{-p}$ decay is
expected (where $p$ is the index of the electron energy distribution). Such
a beaming model would relax the energy requirements on some of the more
extreme GRBs by a factor of several hundreds. It is likely that a large
fraction of the weak (or no) afterglow observations are also due to the
common occurrence of beaming in GRBs, and that their jets have already
transitioned to the spreading phase before the first afterglow observations
were made. With this interpretation, a universal value of $p\cong 2.5$
is consistent with all data.
\end{abstract}

\section{Introduction}

One of the most important open questions in GRBs is whether the burst emission is
isotropic or strongly beamed in our direction. This
question has implications on almost every aspect of the phenomenon, from the
energetics of the events, to the engineering of the ``inner engine'' to the
statistics and the luminosity function of the sources. We suggest here that
GRB~980519 was a jet with an opening angle of less than $0.1$ rad. We also
suggests that such jets are common in most GRBs.

According to the relativistic fireball model, the emission from a
spherically expanding shell and a jet would be rather similar to each other
as long as we are along the jet's axis and the Lorentz factor, $\gamma$, is
large compared to the inverse of the angular width of the jet, $\theta _{0}$
(Piran, 1995). When $\gamma $ drops below $\theta _{0}^{-1}$ the jet's
material begin to spread sideways and we expect a break in the light curve
of the afterglow at this stage. Since we have for spherical adiabatic
evolution $\gamma (t)\approx 6 (E_{52}/n_{1})^{1/8}t_{{\rm day}}^{-3/8}$,
this break should take place at\footnote{
The following numerical factors are different from those given by Rhoads
(1998) and Panaitescu, \& M\'{e}sz\'ros (1998). We explain these differences
in section 2.} 
\begin{equation}  \label{t_jet}
t_{{\rm jet}}\approx 6.2 (E_{52}/n_{1})^{1/3}(\theta
_{0}/0.1)^{8/3} {\rm hr},
\end{equation}
where $E_{52}$ is the ``isotropic'' energy of the ejecta in units of $10^{52}
$ergs, i.e., the inferred energy assuming isotropic expansion, and $n_1$ is
the surrounding ISM particle density in cm$^{-3}$. So far, with the
exception of the recent GRB~990123 (Kulkarni et al. 1999), 
no such break was observed, even for afterglows extending for hundred
days. More specifically, the best observed afterglows, those of GRB~970228
and GRB~970508, behave according to a single unbroken power law, as long as
the observations continued (Zharikov, Sokolov, \& Baryshev 1998; Fruchter et
al. 1998), giving a strong indications that those sources were isotropic to
a large extent.

We show here that, even without seeing a break in the lightcurve, one can
identify a jet based on the powerlaw index of the optical light curve
decline. Since we have a reasonable knowledge of the value of the electrons'
energy distribution index $p\sim 2.4$ we expect for high
frequencies a spherical decay of $t^{-1.1-1.3}$ and a jet like decay of $%
t^{-2.4}$. We suggest that at least in one afterglow, GRB~980519, the
observed light curve and spectra are consistent with an expanding jet and
inconsistent with those expected from a spherical expansion. We suggest that
in this burst the transition to spreading jet, at $\gamma \sim \theta_{0}^{-1}$, 
took place during the few hours between the GRB observations and
the first detection of the afterglow. We conclude that the beaming factor in
this burst is at least a few hundred. 
Together with the appearance of a sharp break
in the light curve of the afterglow of GRB~990123, this indicates that jets
are common in GRBs. In fact the rapid decline that corresponds to an
expanding jet could also explain the weak or no optical afterglow seen in
some of the other bursts, e.g., GRB~990217 (Piro et al. 1999; Palazzi et al.
1999).

Jets have been discussed extensively in the context of GRBs. First the
similarity between some of the observed features of blazars and AGNs led to
the speculation that jets appears also in GRBs (Paczynski 1993; Dermer \&
Chiang 1998). Second, the regions emitting the GRB as well as the afterglow
must be moving relativistically. The emitted radiation is strongly beamed
and we can observe only a region with an opening angle $1/\gamma $ off the
line of sight. Emission outside of this very narrow cone is not
observed. This have lead to numerous speculations on the existence of jets
and to attempts to search for observational signature of jets both during
the GRB phase (Mao and Yi 1994) and in the context of the
afterglow (Rhoads, 1997a,b;1998; M\'{e}sz\'{a}ros et al., 1998; Panaitescu
\& M\'{e}sz\'{a}ros 1998). Finally, GRBs appear naturally in the context of
several leading scenarios for the ''inner engine''. (Mochkovich et al.,
1993; Davies et al., 1994; Katz, 1997, M\'{e}sz\'{a}ros \& Rees, 1997, Nakamura, 1997).

\section{Jet Evolution}

The simple fireball model [and the Blandford-McKee (1977) solution]
assumes a spherical expansion. However, even if the relativistic
ejecta is beamed then as long as the Lorentz factor $\gamma $ of the
relativistic motion satisfies $\gamma >\theta _{0}^{-1}$, the
hydrodynamics of the jet won't be influenced by the fact that it has a
finite angular size (Piran, 1995). The matter doesn't have enough time
(in its own rest frame) to expand sideways.  This situation changes
drastically when $\gamma \approx \theta _{0}^{-1}$ when the sideway
expansion becomes significant. A full solution of the evolution of a
jet at this stage requires 2D relativistic hydro simulations. However,
one can obtain a reasonable idea on what goes on using simple
analytical estimates.

Rhoads (1997a,b;1998) considered the evolution of a relativistic jet that
is expanding sideways at the local speed of sound, $c_{s}$ so that $\theta
\sim \theta _{0}+c_{s}t_{proper}/ct\sim \theta _{0}+\gamma ^{-1}/\sqrt{3}$. 
In this case the hydrodynamic transition takes place at $\gamma \sim
\theta^{-1}/\sqrt{3}$.  However, as the rest mass of the shocked
material is negligible compared with its internal energy, the
expansion can be ultra relativistic with a Lorentz factor comparable
to the thermal Lorentz factor.  This would lead to $\theta \sim \theta
_{0}+ct_{proper}/ct\sim \theta _{0}+\gamma ^{-1}$ and to a transition 
when 
$\gamma \sim \theta _{0}^{-1}$. 
The sideways expansion leads, for an adiabatic evolution, to an exponential
slowing down as $\gamma \propto \exp {[-r/l_{jet}]}$, where $l_{jet}\equiv
[E_{jet}/(4\pi /3)nm_{p}c^{2}]^{1/3}$ is the Sedov length in which a
spherical expanding shell with energy $E_{jet}$ acquire mass whose rest mass
energy equals to its own energy ($n$ is the ISM density). $E_{jet}$ is the
actual energy in the jet. Thus, $r$ is practically a constant during the
spreading phase. Therefore,
the observer time, which is related to the radius and the Lorentz factor as $%
t\propto r/\gamma ^{2}$ satisfies simply $t\propto \gamma ^{-2}$.

Our estimate for the break time (Equation \ref{t_jet}) is the simplest
one.  It is based just on spherical adiabatic expansion. It differs by
a factor of 20 in time (corresponding to a factor of $\sim 3$ in the opening
angle $\theta _{0}$) from the expression given by Rhoads (1998). 
The discrepancy arises from several factors:
(i) As discussed above, we assume that the jet expands sideways
at the speed of light while Rhoads (1998) assumes that jet expands 
at the sound speed $c/\sqrt{3}$.
(ii) Rhoads (1998) uses $t=R/2\gamma ^{2}c$. This expression is valid
for a point source moving along the line of sight with a constant velocity.
We use $t\approx R/4\gamma ^{2}c$ reflecting the deceleration of the source
and its finite angular size (Sari, 1997,1998; Waxman 1997, Panaitescu
and M\'{e}sz\'{a}ros  1998).
(iii) We use the simple adiabatic energy condition: $E=\gamma ^{2}mc^{2}$
, where $m$ is the rest mass of the shocked ISM, while Rhoads (1998) uses $
E=2\gamma ^{2}mc^{2}$. A third possibility is to use the more exact
numerical factor derived from the Blandford McKee (1976) solution: $
E=12\gamma ^{2}mc^{2}/17$.
(iv) We estimated the time in the local frame as $R/\gamma c$.
Rhoads noted that the Lorentz factor was higher earlier and hence the effective
proper time is shorter by a factor of 2.5 allowing for less spreading.
However, far from the shock, the matter moves with a considerably lower 
Lorentz factor allowing it to spread more easily. 

Panaitescu \& M\'{e}sz\'{a}ros (1998) consider similar hydrodynamics as
Rhoads (1998) but notice that once $\gamma \sim 1/\theta _{0}$ the
observer is able to see the edge of the jet. They find two transitions, the first 
one when $\gamma \sim 1/\theta _{0}$ at around our break time estimate and the second
one around Rhoads'. However, there would be only one transition if the time between 
the two breaks turns out to be very short. A reliable estimate of the numerical
factor clearly requires full 2D simulations. It might also be, as suggested
by Rhoads(1998b), that the transition takes place over a relatively long
time and that most observations, that are conducted in a finite time
interval, will show only part of the asymptotic break.

We consider now synchrotron emission from a powerlaw distribution of
accelerated electrons produced by shocks in an expanding jet. The
instantaneous  spectrum is given by the four broken power laws discussed in
Sari, Piran \& Narayan (1998). However, the time dependence of the break
frequency and the overall normalization depend strongly on the hydrodynamic
evolution. Therefore, the lightcurve from a jet differs strongly from the
light curve of a spherical evolution. Surprisingly, it is possible to obtain
general expressions, appropriate to both spherical and jet evolution (by
spherical we mean any system with $\gamma >\theta ^{-1}$ and by a jet a
system with $\gamma \le \theta ^{-1}$). We write these generalized
expressions and specialize to jet and sphere only at the very end. We begin
with the typical frequency $\nu _{m}$, at the observer frame: 
\begin{equation}
\nu _{m}=\frac{eB}{m_{e}c}\gamma _{e}^{2}\gamma \propto 
\gamma ^{4}\propto \cases{
t^{-3/2} & spherical, \cr t^{-2} & jet.}
\end{equation}
The cooling frequency, the synchrotron frequency of electrons that cool on
the dynamical time of the system, is given by 
\begin{equation}
\nu _{c}=\frac{36\pi ^{2}em_{e}c}{\gamma B^{3}t^{2}}\propto \gamma ^{-4}t^{-2}
\propto 
\cases{ t^{-1/2} & spherical, \cr {\rm const.} & jet.}
\end{equation}
The peak flux is obtained at the lowest of the two frequencies $\nu _{m}$
and $\nu _{c}$. Let $\bar{N}_{e}$ be the total number of electrons radiating
towards the observer, {\it i.e.}, those located in a cone of opening angle $%
\gamma ^{-1}$. [$\bar{N}_{e}$ is different from $N_{e}$ (Sari et al., 1998;
Sari \& Piran, 1999) which is the total number of radiating electrons,
including those that are not radiating towards the observer]. $\bar{N}_{e}$
can be approximated by $\bar{N}_{e}=\pi \gamma ^{-2}R^{3}n/3$. The total
energy per unit time per unit frequency emitted by these electrons, $\sigma
_{T}m_{e}c^{2}\bar{N}_{e}B\gamma /6\pi e$, is distributed over an area of $%
\pi \gamma ^{-2}d^{2}$ at a distance $d$ from the source. The observed peak
flux density is therefore 
\begin{equation}
F_{\nu ,max}=\frac{2\sigma _{T}m_{e}c^{2}}{\pi e}\frac{R^{3}nB\gamma }{d^2}
\propto R^{3}\gamma ^{2}\propto \cases{ {\rm const.} & spherical, \cr t^{-1} &
jet.}
\end{equation}

It seems to hold quite generally at late times (except perhaps the first few
hours, see Sari \& Piran, 1999) that $\nu _{c}\gg \nu _{m}.$ The electrons
responsible for low energy emission are therefore those with $\nu _{m}$. In
this case, the self absorption frequency can be estimated as 
\begin{equation}
\nu _{a} \propto R^{3/5}\gamma ^{2/5} \propto 
\cases{ {\rm const.} & spherical \cr t^{-1/5} & jet}.
\end{equation}

We now turn to calculate the light curves for several frequency ranges. The
flux at low frequencies, which is self absorbed, evolves as 
\begin{equation}
F_{\nu <\nu _{a}}\propto R^{2}\propto \cases{t^{1/2} & spherical, \cr {\rm const.}
& jet.}
\end{equation}
The flux at frequencies that are above the self absorption frequency but
below the typical frequency $\nu _{m}$ evolve as 
\begin{equation}
F_{\nu _{a}<\nu <\nu _{m}}\propto R^{3}\gamma ^{2/3}\propto \cases{ t^{1/2} &
spherical, \cr t^{-1/3} & jet.}
\end{equation}
We therefore expect that the low frequencies ($\nu <\nu _{m}$) flux would
rise like $t^{1/2}$ as long as the evolution is spherical. Then, once $%
\gamma $ drops below $\theta _{0}^{-1}$ and the jet begins to spread, the
flux at frequencies above the self absorption would decrease as $t^{-1/3}$.
At lower frequencies which are in the self absorbed regime the flux will be
a constant until the self absorption frequency is reached.

These predictions are different from those derived by Rhoads, who considered
the case when $\nu _{m}<\nu _{a}$ where he found that the flux rises linearly
with time. However, based on GRB~970508, it seems that this regime of $\nu
_{m}<\nu _{a}$ is relevant only for very late times, about a hundred days
after the burst.

At high frequencies two light curves are possible, depending whether the
radiating electrons are cooling ($\nu >\nu _{c}$) or not ($\nu <\nu _{c}$).
The slope itself also depends on the electron power low distribution index $p
$. Below the cooling frequency we obtain 
\begin{equation}
F_{\nu _{m}<\nu <\nu _{c}}=F_{\nu _{m}}\left( \nu /\nu _{m}\right)
^{-(p-1)/2}  \label{numnuc}
\end{equation}
\[
\propto R^{3}\gamma ^{2p}\propto 
\cases{t^{-3(p-1)/4} & spherical, \cr t^{-p} & jet.}
\]
Above the cooling frequency we have 
\begin{equation}
F_{\nu _{m}<\nu _{c}<\nu }=F_{\nu _{m}}\left( \nu _{c}/\nu _{m}\right)
^{-(p-1)/2}\left( \nu /\nu _{c}\right) ^{-p/2}\propto   \label{nucnum}
\end{equation}
\[
R^{3}\gamma ^{2p-2}t^{-1} \propto
\cases{t^{-3p/4+1/2} & spherical,  \cr t^{-p} &  jet.}
\]
Note that for a spreading jet, the light curve decay index 
(but not the spectrum) is independent of
whether $\nu >\nu _{c}$ or $\nu <\nu _{c}$. This is due to the fact that $%
\nu _{c}$ is constant in time in the case of a spreading jet. Since $p$
determines both the light curve and spectrum, a parameter free relation
between the temporal decay index $\alpha $ and the spectral index $\beta $
can be given. For a spherical expansion we have: 
\begin{equation}
\alpha =\cases { 3 \beta/2 & $ \nu < \nu_c$,
\cr 3 \beta/2 -1/2  & $\nu > \nu_c$.}
\label{beta_spherical}
\end{equation}
While for an expanding jet we have: 
\begin{equation}
\alpha =\cases { 2 \beta +1 & $ \nu < \nu_c$,
\cr 2 \beta  & $\nu > \nu_c$.}  
\label{beta_jet}
\end{equation}
These results are summarized in Table 1.

\begin{center}
\begin{table*}[ht!]
\begin{center}
\begin{tabular}{|c||c||c|c|}
\hline
& spectral index & \multicolumn{2}{|c|}{light curve index $\alpha$, $F_{\nu}\propto
t^{-\alpha}$} \\ 
& $\beta$, $F_{\nu}\propto \nu^{-\beta}$ & sphere & jet \\ \hline\hline
&  & $\alpha=3(p-1)/4\cong 1.05$ & $\alpha=p\cong 2.4$ \\ 
\raisebox{1.5ex}[0pt]{$\nu<\nu_c$} & \raisebox{1.5ex}[0pt]{$(p-1)/2
\cong0.7$} & $\alpha=3\beta/2$ & $\alpha=2\beta+1$ \\ \hline
&  & $\alpha=(3p-2)/4\cong 1.3$ & $\alpha=p\cong 2.4$ \\ 
\raisebox{1.5ex}[0pt]{$\nu>\nu_c$} & \raisebox{1.5ex}[0pt]{$ p/2
\cong1.2$} & $\alpha=3\beta/2-1/2$ & $\alpha=2\beta$ \\ \hline
\end{tabular}
\end{center}
\par
\label{t:afterglow}
\caption{The spectral index $\beta$ and the light curve index $\alpha$ as
function of $p$. Typical values are quoted using $p=2.4$. The parameter free
relation between $\alpha$ and $\beta$ is given for each case (eliminating $p$%
). }
\end{table*}
\end{center}

\section{Observations}

\subsection{GRB~980519}

GRB~980519 was one of the brightest of the bursts detected in the BeppoSAX
WFC (Muller et al. 1998; in 't Zand et al. 1999), second only to the recent
GRB~ 990123 (Feroci et al. 1998). The BATSE measured fluence above 25~keV
was $(2.54\pm 0.41)\times 10^{-5}$ ergs~cm$^{-2}$, which places it among the
top 12\% of BATSE bursts (Connaughton 1998). An X-ray observation with the
BeppoSAX Narrow Field Instruments detected an afterglow (Nicastro et al.
1998). GRB~980519 had the most rapid fading of the well-documented GRB
afterglows, consistent with $t^{-2.05\pm 0.04}$ in $BVRI$ (Halpern et al.
1999). The power-law decay index of the X-ray afterglow, $\alpha _{{\rm x}%
}=2.07\pm 0.11$ as reported by Owens et al. (1998), is consistent with the
optical. The X-ray temporal decay of GRB~980518 is the fastest of the seven
afterglows that were well measured by BeppoSAX (Owens et al. 1998). The
optical spectrum alone is well fitted by a power law of the form $\nu
^{-1.20\pm 0.25}$, while the optical and X-ray spectra together are
adequately fitted by a single power law $\nu ^{-1.05\pm 0.10}$.

The relation between the spectral slope and the temporal decay is
inconsistent with the simple spherical fireball model that predicts that the
time decay light curve index, $\alpha $, and the spectral shape power law
index, $\beta $, are related according to equation \ref{beta_spherical}.
This inconsistency is independent of the value of $p$. These observations
are consistent with each other if we assume an expanding jet phase for which
equation \ref{beta_jet} is applicable.


It is difficult to determine the exact value of $p$ from these
observations.  However, we note that they are consistent with a value
of $p\sim 2.4$ that arises in other bursts.  This will fit the optical
power law decay if the electrons are not cooling i.e., $\nu _{c}$ is
above the optical band. It will also fit the optical spectral index
which has large uncertainty. The optical to X-rays slope is intermediate
between the value obtained for slow cooling ($\sim -0.8)$  and
that obtained for fast cooling (-1.25). This indicates that the cooling
frequency is between the optical and x-rays.

The fact that this transition took place less than 8.5 hours after the burst
shows that the opening angle of this jet was rather small: $\theta <0.1$,
leading to a beaming factor of 300 or larger!
We note that the two strongest GRBs detected by the BeppoSAX WFC are inferred
to have a large beaming factor. This may indicate that a significant
fraction of the spread in luminosities is given by the beaming effect.

\subsection{GRB~990123}

GRB~990123 was a remarkable burst with a very high GRB fluence and with a
prompt optical and X-ray emission. We interpret this emission as resulting
from the early forward shock (X-ray) and from the early reverse shock
(optical). The reverse shock has also produced the early radio flare
(Sari and Piran 1999). The forward shock emission is directly related to
the later optical and X-ray emission, while the reverse shock emission
decayed like $t^{-2}$, and disappeared quickly. The late optical
afterglow showed a power law decay with $t^{-1.1\pm 0.03}$. This
behavior continued from the first late observation (about 3.5 hours after
the burst) until about $2.04\pm 0.46$ days after the burst. Then the optical
emission began to decline faster (Kulkarni et al. 1999). The simplest
explanation is that we have observed the transition from a spherical like
phase to an expanding jet phase. The transition took place at $\sim 2$ days,
corresponding to $\theta _{0}\sim 0.1$. This implies a beaming factor of
about $100$ reducing the energy of the burst to $3\times 10^{52}$ ergs.
This is the first, and by now the only, burst in which such a break was
detected. The decay before the break is well measured and fits an electron
distribution with $p \cong 2.5.$

\subsection{GRB~980326}

GRB~980326 was another burst with a rapid decline. Groot et al. (1998)
derived a temporal decay slope of $\alpha =2.1\pm 0.13$ and a spectral slope
of $\beta =0.66\pm 0.7$ in the optical band. Such rapid temporal
decay suggests a jet like evolution. As Groot et al. (1998) note,
the large uncertainty in the spectral index allows in this case also a
spherical expansion interpretation (with somewhat unusual values $p=4.2$ or $p=5.2$). 
However, this measured temporal decay was
dependent upon a report of a host galaxy detection at $R=25.5\pm 0.5$, which
was included as a constant term. The detection of a host has since been
determined to be spurious; better data show no constant component to a
limiting magnitude of $R=27.3$ (Bloom \& Kulkarni 1998). 
When the previously assumed constant component is removed,
the overall light curve
is concave, in disagreement with a jet interpretation. If the last detection is
interpreted as a different phenomenon (Kulkarni, 1999) then the remaining points
show a rapid decline - in agreement with a jet.


\subsection{GRB~970228 and GRB~970508}

In both GRB970228 and GRB970508 there was no observed break in the light
curve as long as the afterglow could be observed. GRB970228 was observed by 
{\it HST} six months later, at which point it was still following a
power-law decay as $t^{-1.14 \pm 0.05}$ (Fruchter et al. 1998). GRB970508
was observed for 9 months to decline as $t^{-1.23 \pm 0.04}$ (Zharikov et al.,
1998), at which point became as faint as its host galaxy. This set a limit
on the beaming in these events of $\theta _{0}\ge 1$. The beaming factor is
therefore less than an order of magnitude.

\section{Conclusions}

We have seen indication of a jet-like behavior in three bursts. Two other
bursts did not show any break in their optical light
curves which have been observed for a long time. In several other bursts the
situation is inconclusive and their short afterglow is consistent with
rather narrow jets. We suggest that jet-like behavior is the common one in
GRBs. Moreover, the range of possible beaming angles, from $\theta _{0}\le
0.1$ for GRB~980519, to $\theta \sim 0.1$ for GRB~990123 to $\theta \ge 1$
for GRB~970228 and GRB~970508 is quite large. These beaming angles are
consistent with the limits set by searches for ``Orphan'' radio (Perna
and Loeb, 1998) and X-ray (Grindlay 1999) afterglows.

The suggestion that GRBs are beamed has several implications. First,
this implies that the GRB ``inner engines'' must include a collimation
mechanism in addition to the required acceleration mechanism. This
makes the similarity between GRBs and some AGNs, more specifically
Blazars, even greater. Second the beaming reduces the energy budget of
this phenomenon.  Beaming of $\theta_0 \sim 0.1$ reduces the required
energy by a factor of 400. Interestingly, the evidence for jets arises
most clearly in the two strongest bursts detected by BeppoSAX so
far. It may provide a hint on the energy budget and on the effect of beaming
on the luminosity function.  GRB models based on
``regular'' compact objects become more appealing once more.

What would be the effects of beaming on the observed luminosity function of
GRBs and on possible intrinsic correlations between different features of
the source? One might expect that different viewing angles might have a very
strong effect on the observed luminosities and other characteristics of the
GRBs. However, as implied by the compactness problem typical initial Lorentz
factors during the GRB, $\gamma_0 \ge 100$. An observer that is more than $%
\gamma_0^{-1}$ away from a GRB jet will practically miss the GRB. Observers,
within this range will see a weaker, softer and longer GRB. However, Mao and
Yi (1994) point out that if $\theta_0 \gg \gamma_0^{-1}$ then the fraction
of bursts that are viewed ``sideways'' is rather small. It would be
approximately $\gamma_0^{-1} / \theta_0$, a few percent, if we use the value
inferred for GRB~990123. Thus this will introduce a small population
of intrinsically weaker, softer and longer bursts. Since these bursts will be
weaker, we expect that we would observe only a small fraction of them.

\acknowledgments
This research was supported by the US-Israel BSF grant 95-328, by the
Israeli Space Agency, by NASA grant NAG5-3516 and by the Sherman Fairchild
Foundation.

\end{document}